\documentclass[preprint,prc,groupedaddress,amsfonts,11pt]{revtex4}
\usepackage{amssymb}
\usepackage{dsfont}
\usepackage{amsmath}

\renewcommand{\arraystretch}{0.8}

\DeclareMathOperator{\Tr}{Tr}

\begin{document}

\title{Noncanonicaly Embedded Rational Map Soliton in Quantum SU(3) Skyrme
Model}
\author{D. Jur\v ciukonis}
\email{darius@itpa.lt}
\author{E.Norvai\v sas}
\email{norvaisas@itpa.lt}
\affiliation{Vilnius University, Institute of Theoretical Physics and Astronomy,\\
Go\v stauto 12, Vilnius 01108, Lithuania}

\begin{abstract}
The quantum Skyrme model is considered in non canonical bases
$\mathrm{SU(3)}\supset \mathrm{SO(3)}$ for the state vectors. A
rational map ansatz is used to describe the soliton with the
topological number bigger than one. The canonical quantization of
the Lagrangian generates in Hamiltonian five different "moments of
inertia" and negative quantum mass corrections, which can
stabilize the quantum soliton solution. Explicit expressions of
the quantum Lagrangian and the Hamiltonian are derived for this
model soliton.
\end{abstract}

\maketitle

\section{Introduction}

The Skyrme model was introduced as an effective theory of baryons
\cite{Skyrme}. Recently the topological soliton solutions namely
the skyrmions, are applied in studies of the quantum Hall effect,
Bose-Einstein condensate and black hole physics. General
analitical solutions of the model are unknown even in the
classical case, and approximate solitonic solutions are under
intensive consideration. For instance, the rational map ansatz
\cite{Manton} is used to describe light nuclei as quantized
skyrmions \cite{Matuzas,Manko}.

The original Skyrme model was defined for a unitary field
$U(\mathbf{x,}t)$ that belongs to a fundamental representation of
the SU(2) group. A semiclassical quantization suggests that the
skyrmion rotates as a "rigid body" \cite{Adkins}. The collective
coordinates approach separates variables which depend on the time
and spatial coordinates. The structure of the ansatz which depends
on spatial coordinates determines the solitonic solutions. A
constructive realization of the canonical quantization adds a term
in the Hamiltonian which may be interpreted as an effective pion
mass term \cite{Acus}. An extension of the model to the SU(N)
group \cite{Walliser} represents a common structure of the Skyrme
Lagrangian and explains its wide application.

The aim of this work is to discuss the group-theoretical aspects
of the canonical quantization of the SU(3) Skyrme model in the
rational map antsatz approximation with baryon number $B\geq 2$.
The ansatz is defined in noncanonical $\mathrm{SU(3)}\supset
\mathrm{SO(3)}$ bases as an SO(3) solitonic solution. The
canonical quantization generates five different "moments of
inertia". The proposed ansatz can be used to describe light nuclei
as special skyrmions.

The paper is organized as follows. Some preliminary definitions
are presented in Sec. 2. The quantum Skyrme model is constructed
\textit{ab initio} in the collective coordinates framework for the
rational map soliton in Sec. 3. Section 4 contains a summarizing
discussion. A few relevant mathematical details and long
expressions of the Hamiltonian terms are given in the Appendix.

\section{Noncanonical embedding of the rational map soliton}

The Skyrme model is a Lagrangian density for a unitary field
$U(\mathbf{x,}t)$ that belongs to the general representation of
the SU(3) group \cite{Jurciukonis}. We consider the unitary field
in the fundamental representation (1,0) of the SU(3) group. The
chiraly symmetric Lagrangian density has the form:
\begin{equation}
{\mathcal{L}}=-\frac{f_{\pi
}^{2}}{4}\mathrm{Tr}\Bigl(\mathbf{R}_{\mu }\mathbf{R}^{\mu
}\Bigr)+\frac{1}{32\mathrm{e}^{2}}\mathrm{Tr}\Bigl([\mathbf{R}_{\mu
},\mathbf{R}_{\nu }][\mathbf{R}^{\mu },\mathbf{R}^{\nu }]\Bigr),
\label{a1a}
\end{equation}%
where the "right" and "left" chiral currents are defined as
\begin{eqnarray}
\mathbf{R}_{\mu } &=&\left( \partial _{\mu }U\right)
\overset{+}{U}=\partial _{\mu }\alpha ^{i}C_{i}^{(B)}(\alpha
)\left\langle \hspace{0.2cm}\left\vert
\hspace{0.1cm}J_{(B)}\hspace{0.05cm}\right\vert
\hspace{0.2cm}\right\rangle ,
\label{a1b} \\
\mathbf{L}_{\mu } &=&\overset{+}{U}\left( \partial _{\mu }U\right)
=\partial _{\mu }\alpha ^{i}C_{i}^{\prime (B)}(\alpha
)\left\langle \hspace{0.2cm}\left\vert
\hspace{0.1cm}J_{(B)}\hspace{0.05cm}\right\vert
\hspace{0.2cm}\right\rangle , \label{a1c}
\end{eqnarray}%
and have the values on the SU(3) algebra. The $f_{\pi }$ and
$\mathrm{e}$ in (\ref{a1a}) are phenomenological parameters of the
model. Explicit expressions of functions $C_{i}^{(B)}(\alpha )$
and $C_{i}^{\prime (B)}(\alpha )$ depend on the group
parametrization $\alpha ^{i}$. The noncanonical SU(3) generators
may be expressed in terms of the canonical generators
$J_{(Z,I,M)}^{(1,1)}$ defined in \cite{Jurciukonis07}:
\begin{eqnarray}
J_{(1,1)}=\sqrt{2}\left(
J_{(\frac{1}{2},\frac{1}{2},\frac{1}{2})}^{(1,1)}-J_{(-\frac{1}{2},\frac{1}{2},\frac{1}{2})}^{(1,1)}\right)
\,,
&\qquad J_{(1,0)}=&2J_{(0,1,0)}^{(1,1)}\,,  \notag \\
J_{(1,-1)}=\sqrt{2}\left(
J_{(-\frac{1}{2},\frac{1}{2},-\frac{1}{2})}^{(1,1)}+J_{(\frac{1}{2},\frac{1}{2},-\frac{1}{2})}^{(1,1)}\right)
\,,
&\qquad J_{(2,2)}=&-2J_{(0,1,1)}^{(1,1)}\,,  \notag \\
J_{(2,1)}=-\sqrt{2}\left(
J_{(\frac{1}{2},\frac{1}{2},\frac{1}{2})}^{(1,1)}+J_{(-\frac{1}{2},\frac{1}{2},\frac{1}{2})}^{(1,1)}\right)
\,,
&\qquad J_{(2,0)}=&-2J_{(0,0,0)}^{(1,1)}\,,  \notag \\
J_{(2,-1)}=-\sqrt{2}\left(
J_{(-\frac{1}{2},\frac{1}{2},-\frac{1}{2})}^{(1,1)}-J_{(\frac{1}{2},\frac{1}{2},-\frac{1}{2})}^{(1,1)}\right)
\,, &\qquad J_{(2,-2)}=&2J_{(0,1,-1)}^{(1,1)}.  \label{a5}
\end{eqnarray}%
The canonical generators satisfy these commutation relations
\begin{equation}
\left[ J_{(L^{\prime },M^{\prime })},J_{(L^{\prime \prime
},M^{\prime \prime })}\right] =-2\sqrt{3}\left[
\renewcommand{\arraystretch}{0.8}\begin{array}{ccc}
\scriptstyle(1,1) & \scriptstyle(1,1) & \scriptstyle(1,1)_{a} \\
L^{\prime } & L^{\prime \prime } & L\end{array}\right] \left[
\begin{array}{ccc}
L^{\prime } & L^{\prime \prime } & L \\
M^{\prime } & M^{\prime \prime } & M\end{array}\right] J_{(L,M)}.
\label{6}
\end{equation}
The bracketed coefficients on the rhs of (\ref{6}) are the SU(3)
noncanonical isofactor and the SO(3) Clebsch-Gordan coefficient.
The state vectors for the canonical bases $\mathrm{SU(3)}\supset
\mathrm{SU(2)}$\ and the non canonical bases
$\mathrm{SU(3)}\supset \mathrm{SO(3)}$\ are equivalent in
fundamental representation. We suggest to use the rational map
antsatz in the SO(3)\ case as a matrix
\begin{eqnarray}
\left( U_{R}\right) _{M,M^{\prime }} &=&D_{M,M^{\prime}}^{1}(\varkappa)=\left( \exp (2i \hat{n}_{a}J_{(1,a)}F(r)) \right)_{M,M^{\prime }} \notag \\
&=& 2\sin ^{2}F(-1)^{M}\hat{n}_{-M}\hat{n}_{M^{\prime }}
 +i\sqrt{2}\sin 2F\left[
\begin{array}{ccc}
1 & 1 & 1 \\
M & u & M^{\prime }\end{array}\right] \hat{n}_{u}+\cos 2F\delta
_{M,M^{\prime }}, \notag \\ \label{rat_antz}
\end{eqnarray}
where the unit vector $\hat{\mathbf{n}}$ is defined \cite{Manton}
in terms of a rational complex function $R(z)$ as
\begin{equation}
\hat{\mathbf{n}}=\frac{1}{1+|R|^{2}}\Bigl\{2\mathop{\mathrm{Re}}(R),
2\mathop{\mathrm{Im}} (R),1-|R|^{2}\Bigr\}\,.  \label{n_vekt}
\end{equation}
The triplet $\varkappa$ of Euler angles is defined by $\hat{n}$
and $F(r)$. By differentiation of $\hat{\mathbf{n}}$ we get an
expression which gives the advantage of the following calculations
\begin{equation}
(-1)^{s}(\nabla _{-s}r\hat{n}_{m})(\nabla _{s}r\hat{n}_{m^{\prime
}})=\hat{n}_{m}\hat{n}_{m^{\prime
}}+\mathcal{I}\left((-1)^{m}\delta _{-m,m^{\prime
}}-\hat{n}_{m}\hat{n}_{m^{\prime }}\right)
\end{equation}
where the symbol $\mathcal{I}$\ denotes the function
\begin{equation}
\mathcal{I}(\theta ,\varphi
)=\left(\frac{1+|z|^{2}}{1+|R|^{2}}\left\vert
\frac{\mathrm{d}R}{\mathrm{d}z}\right\vert \hspace{2pt}
\right)^{2}\,, \label{II}
\end{equation}
that solely depends on the angles $\theta$ and $\varphi$.

The baryon charge density for the rational map skyrmion is expressed as
\begin{equation}
\mathcal{B}(r,\theta ,\varphi )=\varepsilon ^{0k\ell m}\Tr R_{k}R_{\ell
}R_{m}=-\frac{\mathcal{I}\,(\theta ,\varphi )}{2\pi ^{2}}\frac{F^{\prime
}(r)\sin ^{2}F}{r^{2}}.
\end{equation}
Since this expression contains the $\mathcal{I}$ function, there
is no need to modify the usual boundary conditions $F(0)=\pi$ and
$F(\infty)=0$ to account for the chiral angle. The normalized
baryon charge density and the Lagrangian density for SO(3) by a
factor 4. The integral by the spatial angles of $\mathcal{I}$ can
be regarded as the Morse function and is proportional to the
baryon number \cite{Acus06}
\begin{equation}
\int_{0}^{2\pi }\mathrm{d}\varphi \int_{0}^{\pi }\mathrm{d}\theta
\ \mathcal{I}\sin \theta =4\pi B\,.  \label{Idef}
\end{equation}
With the ansatz (\ref{rat_antz}) the Lagrangian density
(\ref{a1a}) reduces to the classical Skyrme Lagrangian for any
baryon number $B$:
\begin{eqnarray}
\mathcal{L}_{\text{cl}}(r,\theta ,\varphi )
&=&-\mathcal{M}_{\mathrm{cl}}=-2f_{\pi }^{2}\left(F^{\prime
2}(r)+\frac{2\mathcal{I}\sin ^{2}F}{r^{2}}\right)  \notag \\
&&-\frac{4}{\mathrm{e}^{2}}\frac{\mathcal{I}\sin
^{2}F}{r^{2}}\left(F^{\prime 2}(r)+\frac{\mathcal{I}\sin
^{2}F}{2r^{2}}\right),  \label{a_Lcl}
\end{eqnarray}
which describes the skyrmion mass density.

\section{Canonical quantization of the Soliton}

The quantization of the model can be carried out by means of
collective coordinates that separate the variables, depending on
the time and spatial coordinates
\begin{equation}
U(\mathbf{r},q(t))=AU_{R}\overset{\phantom{r}+}{A}=A(q(t))U_{R}\left(
\mathbf{r}\right) \overset{\phantom{r}+}{A}(q(t)).  \label{b1}
\end{equation}
Here eight SU(3) group parameters $q^{i}(t),$ $i=1,..8$ are
quantum variables. Thus the Skyrme Lagrangian is considered
quantum mechanically \textit{ab initio} in contrast to the
conventional semiclassical quantization of the soliton as a rigid
body. The generalized coordinates $q^{i}(t)$ and the corresponding
velocities $\dot{q}^{i}\left( t\right) $ satisfy the following
commutation relations
\begin{equation}
\left[ \dot{q}^{\alpha },q^{\beta }\right] =-if^{\alpha \beta }(q),
\label{b2}
\end{equation}
where $f^{\alpha \beta }(q)$ are functions of $q$ only. They are
determined having the quantization condition imposed. The
commutation relation between the velocity component
$\dot{q}^{\alpha }$ and an arbitrary function $G(q)$ is given by
\begin{equation}
\left[ \dot{q}^{\alpha },G(q)\right] =-i\sum_{r}f^{\alpha \beta
}(q)\frac{\partial }{\partial q^{\beta }}G(q).  \label{B3}
\end{equation}
We adopt the usual Weyl ordering for the time derivative:
\begin{equation}
\partial _{0}G(q)=\frac{1}{2}\left\{ \dot{q}^{\alpha },\frac{\partial }{%
\partial q^{\alpha }}G(q)\right\}.  \label{B4}
\end{equation}
The curly brackets denote an anticommutator. Due to the Weyl
operator ordering no further ordering ambiguity appears in the
Lagrangian or the Hamiltonian. The differentiation of the
q-dependent unitary matrix can be expressed in terms of functions
$C_{\alpha }^{\prime (L,M)}$ and the matrix elements of the group
generator $J_{(L,M)}$:
\begin{equation}
\frac{\partial }{\partial q^{\alpha }}G_{(A)(B)}^{(\lambda ,\mu
)}(q)=C_{\alpha }^{\prime (L,M)}(q)G_{(A)(A^{\prime })}^{(\lambda
,\mu )}(q) \genfrac{\langle}{|}{0pt}{}{(\lambda ,\mu )}{A^{\prime
}} J_{(L,M)} \genfrac{|}{\rangle}{0pt}{}{(\lambda ,\mu )}{(B)}
\label{difG}
\end{equation}
The relations (\ref{b2}) -- (\ref{B4}) are neglected in
semiclassical calculations.

For the purpose of defining the metric tensor in the Lagrangian we
use an approximate expression:
\begin{equation}
\overset{\phantom{r}+}{A}\dot{{A}}\thickapprox \frac{1}{2}\left\{
\dot{q}^{\alpha },C_{\alpha }^{\prime (L,M)}(q)\right\}
\left\langle \hspace{0.2cm}\left\vert
\hspace{0.1cm}J_{(L,M)}\hspace{0.05cm}\right\vert
\hspace{0.2cm}\right\rangle +...  \label{AA}
\end{equation}
which will be specified later.

After substitution of the ansatz (\ref{b1}) into the model
Lagrangian density (\ref{a1a}) and integration over spatial
coordinates the Lagrangian has this form:
\begin{equation}
L=\frac{1}{2}\dot{q}^{\alpha }g_{\alpha \beta }(q,F)\dot{q}^{\beta
}+a^{0}\frac{1}{2}\left\{ \dot{q}^{\alpha },C_{\alpha }^{\prime
(2,0)}(q)\right\} +\left[ \left( \dot{q}\right) ^{0}\text{ order
terms}\right],  \label{b22}
\end{equation}
where the metric tensor is
\begin{equation}
g_{\alpha \beta }(q,F)=C_{\alpha }^{\prime (L,M)}(q)E_{(L,M)(L^{\prime
},M^{\prime })}(F)C_{\beta }^{\prime (L^{\prime },M^{\prime })}(q),
\label{b4}
\end{equation}
with the intermediate function $E_{(L,M)(L^{\prime },M^{\prime
})}$ defined as
\begin{equation}
E_{(L,M)(L^{\prime },M^{\prime })}(F)=-(-1)^{M}a_{L,M}(F)\delta
_{L,L^{\prime }}\delta _{M,-M^{\prime }}.  \label{b5}
\end{equation}
Note that the exact expression of the coefficient $a^{0}$ is not
important for the calculation of $g_{\alpha \beta }$. There are
five different quantum moments of inertia in (\ref{b22}):
\begin{subequations}
\begin{align}
a_{(1,0)}(F)& =\frac{1}{\mathrm{e}^{3}f_{\pi }}\int
\mathrm{d}^{3}\tilde{r}\tilde{r}^{2}\sin ^{2}F\left(
n_{0}^{2}-1\right) \left[ 1+F^{\prime 2}+\frac{\mathcal{I}}{r^{2}}\sin ^{2}F\right] ; \\
a_{(1,1)}(F)& =a_{(1,-1)}(F)=\frac{1}{2\mathrm{e}^{3}f_{\pi }}\int
\mathrm{d}^{3}\tilde{r}\tilde{r}^{2}\sin ^{2}F\left(
n_{0}^{2}+1\right) \left[
1+F^{\prime 2}+\frac{\mathcal{I}}{r^{2}}\sin ^{2}F\right] ; \\
a_{(2,0)}(F)& =\frac{1}{\mathrm{e}^{3}f_{\pi }}\int
\mathrm{d}^{3}\tilde{r}\tilde{r}^{2}\sin ^{2}F\left(
n_{0}^{2}-1\right) \biggl(\cos
^{2}F+n_{0}^{2}\sin ^{2}F  \notag \\
& -\left(n_{0}^{2}-4\cos ^{2}F+2n_{0}^{2}\cos 2F\right)F^{\prime 2}  \notag \\
& +\frac{\mathcal{I}}{r^{2}}\sin ^{2}F\left(2\cos
^{2}F+n_{0}^{2}(4-\cos 2F)\right)\biggr); \\
a_{(2,1)}(F)& =a_{(2,-1)}(F)=\frac{1}{2\mathrm{e}^{3}f_{\pi }}\int
\mathrm{d}^{3}\tilde{r}\tilde{r}^{2}\sin ^{2}F\biggl(3+2\cos
2F-3n_{0}^{2}+4n_{0}^{4}\sin ^{2}F  \notag \\
& +\left(9+8\cos 2F-3n_{0}^{2}-4n_{0}^{4}(1+2\cos 2F)\right)F^{\prime 2}  \notag \\
& +\frac{\mathcal{I}}{r^{2}}\sin ^{2}F\left(9+4\cos
2F-15n_{0}^{2}+4n_{0}^{4}(4-\cos 2F)\right)\biggr); \\
a_{(2,2)}(F)& =a_{(2,-2)}(F)=\frac{1}{4\mathrm{e}^{3}f_{\pi }}\int
\mathrm{d}^{3}\tilde{r}\tilde{r}^{2}\sin ^{2}F\biggl( -3-\cos
2F-12n_{0}^{2}\cos
^{2}F+2n_{0}^{4}\sin ^{2}F  \notag \\
& -2\left(3+2\cos 2F-24n_{0}^{2}\cos ^{2}F+n_{0}^{4}(1+2\cos
2F)\right)F^{\prime 2}
\notag \\
& -\frac{2\mathcal{I}}{r^{2}}\sin ^{2}F\left(6+\cos
2F-12n_{0}^{2}\cos ^{2}F-n_{0}^{4}(4-\cos 2F)\right)\biggr);
\end{align}
\end{subequations}
where the dimensionless variable $\tilde{r}=\mathrm{e}f_{\pi }r$
and $\mathrm{d}^{3}\tilde{r}=\sin \theta d\theta d\varphi
d\tilde{r}.$\ These quantum moments depend on the chiral angle
function $F(r)$, one component of the rational map vector $n_{0}$
and the function $\mathcal{I}(\theta ,\varphi )$.

The canonical momenta are defined as
\begin{equation}
p_{\beta }=\frac{\partial L}{\partial \dot{q}^{\beta }}=\frac{1}{2}\left\{
\dot{q}^{\alpha },g_{\alpha \beta }\right\} +a^{0}C_{\beta }^{^{\prime
}(2,0)}(q).  \label{b7}
\end{equation}
Note that the momenta do not commute and have terms which do not
contains velocity. The parametrization $q^{\alpha }$ of the group
manifold is significant for the definition of the canonical
momenta. For the time being we do not require $\left[ p_{\alpha
},p_{\beta }\right] =0$. The momenta and the conjugate coordinates
satisfy the commutation relations $\left[ p_{\beta },q^{\alpha
}\right] =-i\delta _{\alpha \beta }$. This commutation relations
determine the explicit expressions of the functions $f^{\alpha
\beta }(q)$ in (\ref{b2}):
\begin{equation}
f^{\alpha \beta }(q)=\left( g_{\alpha \beta }\right) ^{-1}=C_{(L,M)}^{\prime
\alpha }(q)E^{(L,M)(L^{\prime },M^{\prime })}(F)C_{(L^{\prime },M^{\prime
})}^{\prime \beta }(q),  \label{b8}
\end{equation}
where
\begin{equation}
E^{(L,M)(L^{\prime },M^{\prime })}(F)=-(-1)^{M}\frac{1}{a_{(L,M)}(F)}\delta
_{L,L^{\prime }}\delta _{M,-M^{\prime }}.  \label{b9}
\end{equation}
The functions $C_{(L,M)}^{\prime \alpha }(q)$ are defined in
(\ref{Cinvert}).

It is possible to choose the parametrization on the SU(3) group
manifold so that the eight operators
\begin{equation}
\hat{R}_{(LM)}=\frac{i}{2}\left\{ p_{\beta },C_{(L,M)}^{\prime \beta
}(q)\right\}  \label{b901}
\end{equation}
are defined as the group generators satisfying the commutation
relations (\ref{6}). It is easy to check that because of the
choice (\ref{b901}) the requirement $[p_{\alpha },p_{\beta }]=0$
is satisfied for certain. The proof for the SU(2) group can be
found in \cite{Fujii}. The generators (\ref{b901}) act on the
Wigner matrix of the SU(3) irreducible representation as right
transformation generators:
\begin{equation}
\left[ \hat{R}_{(LM)},D_{(\alpha _{1}L_{1}M_{1})(\alpha
_{2}L_{2}M_{2})}^{(\lambda ,\mu )}(q)\right] =D_{(\alpha
_{1}L_{1}M_{1})(\alpha _{2}L_{2}M_{2})}^{(\lambda ,\mu )}(q)
\genfrac{\langle}{|}{0pt}{}{(\lambda ,\mu )}{(\alpha _{2}^{\prime
}L_{2}^{\prime }M_{2}^{\prime })} J_{(LM)}
\genfrac{|}{\rangle}{0pt}{}{(\lambda ,\mu )}{(\alpha
_{2}L_{2}M_{2})}. \label{b_act}
\end{equation}
The indices $\alpha _{1}$ and $\alpha _{2}$ label the multiplets
of $(L,M)$. For instance, the left transformation generators are
defined as
\begin{equation}
\hat{L}_{(LM)}=\frac{i}{2}\left\{ p_{\beta },C_{(L,M)}^{\beta
}(q)\right\}. \label{b_left}
\end{equation}

Determination of functions $f^{\alpha \beta }(q)$ allows us to
obtain an explicit expression of (\ref{AA}):
\begin{eqnarray}
\overset{\phantom{r}+}{A}\dot{{A}}
&=&\overset{\phantom{r}+}{A}\left\{ \dot{q}^{\alpha },A\right\}  \notag \\
&=&\frac{1}{2}\left\{ \dot{q}^{\alpha },C_{\alpha }^{\prime
(LM)}(q)\right\} \left\langle \hspace{0.2cm}\left\vert
\hspace{0.1cm}J_{(LM)}\hspace{0.05cm}\right\vert
\hspace{0.2cm}\right\rangle
-\frac{i}{2}E^{(L_{1}M_{1})(L_{2}M_{2})}\left\langle
\hspace{0.2cm}\left\vert
\hspace{0.1cm}J_{(L_{1}M_{1})}J_{(L_{2}M_{2})}\hspace{0.05cm}\right\vert
\hspace{0.2cm}\right\rangle  \notag \\
&=&\frac{1}{2}\left\{ \dot{q}^{\alpha },C_{\alpha }^{\prime
(LM)}(q)\right\} \left\langle \hspace{0.2cm}\left\vert
\hspace{0.1cm}J_{(LM)}\hspace{0.05cm}\right\vert
\hspace{0.2cm}\right\rangle
+\frac{i}{2a_{0}}\mathds{1}-\frac{i}{2a_{2}}\left\langle
\hspace{0.2cm}\left\vert
\hspace{0.1cm}J_{(20)}\hspace{0.05cm}\right\vert
\hspace{0.2cm}\right\rangle, \label{b_tiksli_AA}
\end{eqnarray}
where $a_{0}$ and $a_{2}$ are constructed from the quantum moments
of inertia:
\begin{subequations}
\begin{align}
\frac{1}{a_{0}}& =\frac{1}{3}\left(
\frac{2}{a_{(1,0)}}+\frac{4}{a_{(1,1)}}+\frac{2}{a_{(2,0)}}+\frac{4}{a_{(2,1)}}+\frac{4}{a_{(2,2)}}\right) ; \\
\frac{1}{a_{2}}& =\frac{1}{\sqrt{3}}\left(
-\frac{1}{a_{(1,0)}}+\frac{1}{a_{(1,1)}}+\frac{1}{a_{(2,0)}}+\frac{1}{a_{(2,1)}}-\frac{2}{a_{(2,2)}}\right)
.
\end{align}
\end{subequations}
The field (\ref{b1}) is substituted in the Lagrangian density
(\ref{a1a}) in order to obtain the explicit expression in terms of
the collective coordinates and the space coordinates. After long
calculation by using (\ref{b_tiksli_AA}) and the commutation
relation (\ref{B3}) (which is very important), we get a complete
explicit expression of the Skyrme model Lagrangian density
\begin{eqnarray}
\mathcal{L}(q,\dot{q},\varkappa )&=&\Bigl\{\dot{q}^{\alpha
},C_{\alpha }^{\prime (L,M_{1})}(q)\Bigr\}\Bigl\{\dot{q}^{\beta
},C_{\beta }^{\prime
(L,M_{2})}(q)\Bigr\}{\mathcal{V}}_{1}(\varkappa) \notag \\
&&+i\Bigl\{\dot{q}^{\alpha },C_{\alpha }^{\prime
(L,M_{1})}(q)\Bigr\}{\mathcal{V}}_{2}(\varkappa)+{\mathcal{V}}_{3}(\varkappa)-\mathcal{M}_{\mathrm{cl}}.
\label{b_Ld}
\end{eqnarray}
The function ${\mathcal{V}}_{1}$ in first term results from the
trace of two group generators (see (\ref{Tr2}) below). The
function ${\mathcal{V}}_{2}$ results from the trace containing
three group generators (see (\ref{Tr3})). The function
${\mathcal{V}}_{3}$ results from the trace containing four group
generators (\ref{Tr4}). Expressions of ${\mathcal{V}}_{i}$
functions are presented in Appendix, see (\ref{V1}) -- (\ref{V3}).
The terms with functions ${\mathcal{V}}_{2}$ and
${\mathcal{V}}_{3}$ are absent in the semiclassical quantization.

Integration (\ref{b_Ld}) over the space variables gives the
Lagrangian
\begin{equation}
L=\frac{1}{8}\Bigl\{\dot{q}^{\alpha },C_{\alpha }^{\prime
(L_{1},M_{1})}(q)\Bigr\}E_{(L_{1}M_{1})(L_{2}M_{2})}\Bigl\{\dot{q}^{\beta
},C_{\beta }^{\prime (L_{2},M_{2}^{\prime
})}(q)\Bigr\}+i\Bigl\{\dot{q}^{\alpha },C_{\alpha }^{\prime
(2,0)}(q)\Bigr\}V_{2}+V_{3}-M_{\text{cl}}, \label{b_Lag}
\end{equation}
where $V_{i}=\int \mathrm{d}^{3}x{\mathcal{V}}_{i}(\varkappa)$.

R. Sugano and collaborators \cite{Sugano} developed the q-number
variational method to formulate a theory that has the consistency
between the Lagrangian and the Hamiltonian formalisms on the
curved space of generalized coordinates. From (\ref{b_Lag}) we
specify the coefficient $a^{0}=2iV_{2}$ that was undetermined in
(\ref{b7}) and derive the Hamiltonian in a form
\begin{eqnarray}
H &=&\frac{1}{8}\Bigl\{\dot{q}^{\alpha },C_{\alpha }^{\prime
(L_{1},M_{1})}(q)\Bigr\}E_{(L_{1}M_{1})(L_{2}M_{2})}\Bigl\{\dot{q}^{\beta
},C_{\beta }^{\prime (L_{2},M_{2}^{\prime })}(q)\Bigr\}-V_{3}+M_{\text{cl}}
\notag \\
&=&-\frac{1}{2}\hat{R}_{(L_{1}M_{1})}E^{(L_{1}M_{1})(L_{2}M_{2})}
\hat{R}_{(L_{2}M_{2})}-\frac{2V_{2}}{a_{(2,0)}}\hat{R}_{(2,0)}-2\left(
\frac{V_{2}}{a_{(2,0)}}\right) ^{2}-V_{3}+M_{\text{cl}}. \notag \\
\label{b10}
\end{eqnarray}

We define the state vectors as the complex conjugate Wigner matrix
elements of the $(\Lambda ,\Theta )$ representation depending on
eight quantum variables $q^{\alpha }$:
\begin{equation}
\left\vert
\begin{array}{c}
(\Lambda ,\Theta ) \\
\alpha ,S,N;\beta ,S^{\prime },N^{\prime }\end{array}\right\rangle
=\sqrt{\dim (\Lambda ,\Theta )}D_{(\alpha ,S,N)(\beta ,S^{\prime
},N^{\prime })}^{\ast (\Lambda ,\Theta )}(q)\left\vert
0\right\rangle,  \label{c55}
\end{equation}
The indices $\alpha $ and $\beta $ label the multiplets of the
SO(3) group. $\left\vert 0\right\rangle $ denotes the vacuum
state. Because of five different "moments of inertia" the vectors
(\ref{c55}) are not the eigenstates of the Hamiltonian
(\ref{b10}). The action of the Hamiltonian on vectors (\ref{c55})
following (\ref{b_act}) can be expressed in terms of the "moments
of inertia" $a_{(L,M)}$ and the SU(3) group Clebsch-Gordan
coefficients.

We take into account the chiral symmetry breaking effects by
introducing a term
\begin{equation}
\mathcal{M}_{\text{sb}}=\frac{1}{16}f_{\pi
}m_{0}^{2}Tr(U+\overset{+}{U}-2), \label{c6}
\end{equation}
which takes an explicit form
\begin{equation}
\mathcal{M}_{\text{sb}}=\frac{1}{2}f_{\pi }m_{0}^{2}\sin ^{2}F,  \label{c7}
\end{equation}
like in $B=1$ case \cite{Jurciukonis07}.

\section{Conclusion}

We considered a new rational map approximation ansatz for the
Skyrme model which is the noncanonical embedded
$\mathrm{SU(3)}\supset \mathrm{SO(3)}$ soliton with baryon number
$B\geq 2$. The SU(2) rational map ansatz is not spherically
symmetric. The canonical quantization leads to five different
quantum "moments of inertia" in the Hamiltonian and the negative
quantum mass corrections. The state vectors are defined as the
SU(3) group representation $(\Lambda ,\Theta )$ matrix depending
on eight quantum variables $q^{i}$ because the ansatz does not
commute with any generator of the group. The vectors (\ref{c55})
are not eigenvectors of the Hamiltonian for higher
representations. The mixing is small. To find eigenstate vectors,
the Hamiltonian matrix  must be diagonalized in every $(\Lambda
,\Theta )$ representation. If the baryon number $B=1$ and
$\hat{n}=\hat{x}$, we get a soliton with two different "moments of
inertia" which was considered in \cite{Jurciukonis07}.

\appendix
\section{}

\noindent The functions $C_{(L,M)}^{\prime \alpha }(q)$ satisfy
the following orthogonality relations:
\begin{subequations}
\begin{align}
\sum_{LM}C_{\alpha }^{^{\prime }(LM)}(q)\cdot C_{(LM)}^{^{\prime }\beta
}(q)& =\delta _{\alpha \beta },  \label{Cinvert} \\
\sum_{\alpha }C_{\alpha }^{^{\prime }(LM)}(q)\cdot C_{(L^{\prime
}M;)}^{^{\prime }\alpha }(q)& =\delta _{(LM)(L^{\prime }M^{\prime })}.
\end{align}
\end{subequations}
The commutator of the canonical momenta and the functions
$C_{(L,M)}^{\prime \alpha }(q)$ equals to:
\begin{equation}
\frac{1}{2}\left\{ \dot{q}^{\alpha },C_{\alpha }^{\prime (L,M)}(q)\right\}
=E^{(LM)(L^{\prime }M^{\prime })}\frac{1}{2}\left\{ p_{\beta },C_{(L^{\prime
}M^{\prime })}^{\prime \beta }(q)\right\} -2iV_{2}E^{(2,0)(LM)}.
\end{equation}%
The trace of two group generators:
\begin{equation}
\Tr\genfrac{\langle}{|}{0pt}{}{(1,0)}{D}J_{(L_{1},M_{1})}
J_{(L_{2},M_{2})}\genfrac{|}{\rangle}{0pt}{}{(1,0)}{(D)}=(-1)^{M_{1}}
2\delta _{L_{1},L_{2}}\delta _{M_{1},-M_{2}}.  \label{Tr2}
\end{equation}
The trace of three group generators:
\begin{eqnarray}
\Tr\genfrac{\langle}{|}{0pt}{}{(1,0)}{D}
J_{(L_{1}M_{1})}J_{(L_{2}M_{2})}J_{(L_{3}M_{3})}\genfrac{|}{\rangle}{0pt}{}{(1,0)}{(D)}
&=&(-1)^{M_{3}+1}2\sqrt{3}\Biggl(\renewcommand{\arraystretch}{0.9}\left[
\begin{array}{ccc}
\scriptstyle(1,1) & \scriptstyle(1,1) & \scriptstyle(1,1)_{\gamma =1} \\
\scriptstyle(L_{1}M_{1}) & \scriptstyle(L_{2}M_{2}) &
\scriptstyle(L_{3}-M_{3})\end{array}
\right]  \notag \\
&&+\renewcommand{\arraystretch}{0.9}\left[
\begin{array}{ccc}
\scriptstyle(1,1) & \scriptstyle(1,1) & \scriptstyle(1,1)_{\gamma =2} \\
\scriptstyle(L_{1}M_{1}) & \scriptstyle(L_{2}M_{2}) &
\scriptstyle(L_{3}-M_{3})\end{array}\right]
\frac{\sqrt{5}}{3}\Biggl).  \label{Tr3}
\end{eqnarray}
The trace of four group generators:
\begin{eqnarray}
&&\Tr\genfrac{\langle}{|}{0pt}{}{(1,0)}{D}
J_{(L_{1}M_{1})}J_{(L_{2}M_{2})}J_{(L_{3}M_{3})}J_{(L_{4}M_{4})}\genfrac{|}{\rangle}{0pt}{}{(1,0)}{(D)}
=4\left[
(2L_{1}+1)(2L_{2}+1)(2L_{3}+1)\right] ^{\frac{1}{2}}  \notag \\
&&\phantom{}\times \left[(2L_{4}+1)\right] ^{\frac{1}{2}}
\sum_{k}(-1)^{u}\left\{
\begin{array}{ccc}
L_{1} & L_{2} & k \\
1 & 1 & 1\end{array}\right\} \left\{
\begin{array}{ccc}
L_{3} & L_{4} & k \\
1 & 1 & 1\end{array}\right\} \left[
\begin{array}{ccc}
L_{1} & L_{2} & k \\
M_{1} & M_{2} & -u\end{array}\right] \left[
\begin{array}{ccc}
L_{3} & L_{4} & k \\
M_{3} & M_{4} & u\end{array}\right].  \notag \\
\label{Tr4}
\end{eqnarray}
The terms included in the Lagrangian density (\ref{b_Ld}):
\begin{eqnarray}
{\mathcal{V}}_{1}(\varkappa) &=&\frac{f_{\pi
}^{2}}{4}(-1)^{M_{1}}\biggl( D_{-M_{1},M_{2}}^{L}(\varkappa
)+\overset{+\phantom{WWWWii}}{D_{-M,M^{\prime }}^{L}(\varkappa
)}-2\delta_{-M_{1},M_{2}}\biggr)+\frac{1}{16\mathrm{e}^{2}}(-1)^{M^{\prime }}B_{m,m^{\prime }}(\varkappa)   \notag \\
&&\times \frac{3}{5-2L}\left[
\begin{array}{ccc}
L & 1 & L \\
M_{1}^{\prime } & m & M^{\prime }\end{array}\right] \left[
\begin{array}{ccc}
L & 1 & L \\
M_{2}^{\prime } & m^{\prime } & -M^{\prime }\end{array}\right]
\Biggl( 2\delta _{M_{1},M_{1}^{\prime }}D_{M_{2},M_{2}^{\prime
}}^{L}(\varkappa )-\delta _{M_{1},M_{1}^{\prime }}\delta
_{M_{2},M_{2}^{\prime }}  \notag \\
&&-D_{M_{1},M_{1}^{\prime }}^{L}(\varkappa )D_{M_{2},M_{2}^{\prime
}}^{L}(\varkappa )\Biggr) ;  \label{V1}
\end{eqnarray}
\begin{eqnarray}
{\mathcal{V}}_{2}(\varkappa) &=&\frac{f_{\pi
}^{2}}{4}\Biggl(\frac{(-1)^{M_{2}+M_{1}}}{a_{(L,M_{2}})}\frac{\sqrt{2}}{\sqrt{3}}\sqrt{L^{2}+L+1}\left[
\begin{array}{ccc}
L & L & 2 \\
M_{2} & -M_{2}^{\prime } & M_{1}\end{array} \right] \biggl(
D_{M_{2}^{\prime },M_{2}}^{L}(\varkappa )-\overset{+ \phantom{WWWWi}}{D_{M_{2}^{\prime },M_{2}}^{L}(\varkappa )}\biggl)  \notag \\
&&\phantom{}-\frac{1}{a_{2}}\biggl( D_{0,M_{1}}^{2}(\varkappa
)-\overset{+\phantom{WWWi}}{D_{0,M_{1}}^{2}(\varkappa )}\biggr) \Biggr)  \notag \\
&&+\frac{1}{4\mathrm{e}^{2}}(-1)^{M_{1}^{\prime
}+M_{2}}\frac{\sqrt{2\cdot
3}}{a_{(L,M_{2}})}\frac{\sqrt{L^{2}+L+1}}{\sqrt{5-2L}}B_{m,m^{\prime
}}(\varkappa )\left[
\begin{array}{ccc}
2 & 1 & 2 \\
M_{1}^{\prime \prime } & m & M_{1}^{\prime }\end{array}\right]
\left[
\begin{array}{ccc}
L & 1 & L \\
M_{2}^{\prime \prime } & m^{\prime } & M_{2}^{\prime }\end{array}
\right]  \notag \\
&&\times \left[
\begin{array}{ccc}
L & L & 2 \\
M_{2}^{\prime } & M_{2}^{\prime \prime \prime } & -M_{1}^{\prime
}\end{array}\right] \biggl(\delta _{M_{1}M_{1}^{\prime \prime
}}\delta _{M_{2}M_{2}^{\prime \prime
}}\overset{+\phantom{WWWWWii}}{D_{M_{2}^{\prime \prime \prime
},-M_{2}^{\prime }}^{L}(\varkappa )}-\delta _{M_{1}M_{1}^{\prime
\prime }}\delta _{M_{2}M_{2}^{\prime \prime }}\delta
_{-M_{2}M_{2}^{\prime \prime \prime }}  \notag \\
&&\phantom{}-\delta _{M_{1}M_{1}^{\prime \prime }}\delta
_{-M_{2}M_{2}^{\prime \prime \prime }}D_{M_{2}^{\prime \prime
},M_{2}}^{L}(\varkappa )+\delta _{M_{2}M_{2}^{\prime \prime
}}\delta _{-M_{2}M_{2}^{\prime \prime \prime
}}\overset{+\phantom{WWWWii}}{D_{M_{1}^{\prime \prime
},M_{1}}^{2}(\varkappa )}\biggr);  \label{V2}
\end{eqnarray}
\begin{eqnarray}
{\mathcal{V}}_{3}(\varkappa) &=&\frac{f_{\pi
}^{2}}{4}\Bigl(\frac{4(2L_{1}+1)(2L_{2}+1)}{a_{(L_{1},M_{1})}a_{(L_{2},M_{2})}}\left\{
\begin{array}{ccc}
L_{1} & L_{2} & k \\
1 & 1 & 1%
\end{array}
\right\} ^{2}\left[
\begin{array}{ccc}
L_{1} & L_{2} & k \\
M_{1} & M_{2} & u\end{array}\right] ^{2}D_{u,u}^{k}(\varkappa
)+\frac{3}{a_{0}^{2}}  \notag \\
&&\phantom{}+\frac{1}{a_{2}^{2}}\biggl( 1+\overset{+
\phantom{WWii}}{D_{0,0}^{2}(\varkappa
)}\biggr)-\frac{4}{a_{(L,M)}}\Biggl(
\frac{1}{a_{0}}D_{M,M}^{L}(\varkappa
)-(-1)^{M}\frac{1}{a_{2}}\frac{\sqrt{L^{2}+L+1}}{\sqrt{2\cdot
3}}  \notag \\
&&\times\left[
\begin{array}{ccc}
L & L & 2 \\
M & -M & 0\end{array}
\right] D_{M,M}^{L}(\varkappa )\Biggr)  \notag \\
&&-\frac{3}{2\mathrm{e}^{2}}\frac{(2L_{1}+1)(2L_{2}+1)}{\sqrt{\left(
5-2L_{1}\right) \left( 5-2L_{2}\right) }}(-1)^{M_{1}+M_{2}+u}B_{m,m^{\prime
}}(\varkappa )\left\{
\begin{array}{ccc}
L_{1} & L_{1} & k \\
1 & 1 & 1\end{array}\right\} \left\{
\begin{array}{ccc}
L_{2} & L_{2} & k \\
1 & 1 & 1\end{array}\right\}   \notag \\
&&\times \left[
\begin{array}{ccc}
L_{1} & 1 & L_{1} \\
M_{1}^{\prime } & m & M_{1}^{\prime \prime }
\end{array}\right] \left[
\begin{array}{ccc}
L_{2} & 1 & L_{2} \\
M_{2}^{\prime } & m^{\prime } & M_{2}^{\prime \prime }\end{array}
\right] \left[
\begin{array}{ccc}
L_{1} & L_{1} & k \\
M_{1} & -M_{1}^{\prime \prime } & u \end{array} \right] \left[
\begin{array}{ccc}
L_{2} & L_{2} & k \\
-M_{2} & M_{2}^{\prime \prime } & u \end{array}
\right]  \notag \\
&&\times
\Biggl(\frac{1}{a_{(L_{1},M_{1})}a_{(L_{2},M_{2})}}\biggl(\delta
_{M_{2}M_{2}^{\prime }}D_{M_{1}^{\prime },M_{1}}^{L_{1}}(\varkappa
)\left( 1-(-1)^{k}\right) -\delta _{M_{1}M_{1}^{\prime }}\delta
_{M_{2}M_{2}^{\prime
}}  \notag \\
&&\phantom{}-D_{M_{1}^{\prime },M_{1}}^{L_{1}}(\varkappa
)D_{M_{2}^{\prime },M_{2}}^{L_{2}}(\varkappa )\biggr) +
\frac{1}{a_{(L_{1},M_{1}^{\prime })}a_{(L_{2},M_{2})}}\delta
_{M_{2}M_{2}^{\prime }}D_{M_{1}^{\prime },M_{1}}^{L_{1}}(\varkappa
)\left( 1+(-1)^{k}\right) \Biggr).  \notag \\
\label{V3}
\end{eqnarray}
where $B_{m,m^{\prime }}(\varkappa )$ are:
\begin{eqnarray}
B_{m,m^{\prime }}(\varkappa )=8(-1)^{m+m^{\prime
}}\hat{n}_{-m}\hat{n}_{-m^{\prime }}\left(
\frac{1}{r^{2}}\mathcal{I}\sin ^{2}F-F^{\prime 2}\right)
-(-1)^{m}\delta _{m,-m^{\prime }}\frac{8}{r^{2}}\mathcal{I}\sin
^{2}F. \notag \\
\label{Bmm}
\end{eqnarray}

\end{document}